\documentclass[prd,showpacs,preprintnumbers,amsmath,amssymb,superscriptaddress,nofootinbib,english]{revtex4}
\usepackage{graphicx}
\usepackage{amsmath,amssymb}
\usepackage{hyperref}
\usepackage[usenames]{color}
\usepackage{gensymb}

\hypersetup{
    colorlinks=true,
    linkcolor=red,
    citecolor=blue,
}

\def\ie{{\frenchspacing\it i.e.}}
\def\eg{{\frenchspacing\it e.g.}}

\def\be{\begin{equation}}
\def\ee{\end{equation}}
\def\ba{\begin{eqnarray}}
\def\ea{\end{eqnarray}}
\def\nn{\nonumber}

\newcommand\ctbobs{{C_{\ell}^{\rm TB,obs}}}
\newcommand\cteobs{{C_{\ell}^{\rm TE,obs}}}
\newcommand\ceeobs{{C_{\ell}^{\rm EE,obs}}}
\newcommand\cbbobs{{C_{\ell}^{\rm BB,obs}}}
\newcommand\cebobs{{C_{\ell}^{\rm EB,obs}}}

\newcommand\cttrot{{C_{\ell}^{\rm TT,rot}}}
\newcommand\ctbrot{{C_{\ell}^{\rm TB,rot}}}
\newcommand\cterot{{C_{\ell}^{\rm TE,rot}}}
\newcommand\ceerot{{C_{\ell}^{\rm EE,rot}}}
\newcommand\cbbrot{{C_{\ell}^{\rm BB,rot}}}
\newcommand\cebrot{{C_{\ell}^{\rm EB,rot}}}

\newcommand\ctt{{C_{\ell}^{\rm TT}}}
\newcommand\ctb{{C_{\ell}^{\rm TB}}}
\newcommand\cte{{C_{\ell}^{\rm TE}}}
\newcommand\cee{{C_{\ell}^{\rm EE}}}
\newcommand\cbb{{C_{\ell}^{\rm BB}}}
\newcommand\ceb{{C_{\ell}^{\rm EB}}}

\newcommand\ctbp{{C_{\ell'}^{\rm TB}}}
\newcommand\ctep{{C_{\ell'}^{\rm TE}}}
\newcommand\ceep{{C_{\ell'}^{\rm EE}}}
\newcommand\cbbp{{C_{\ell'}^{\rm BB}}}
\newcommand\cebp{{C_{\ell'}^{\rm EB}}}
\newcommand\FXX{{\bf F_{\Delta\Delta}}}
\newcommand\FYY{{\bf F_{\Xi\Xi}}}
\newcommand\FXY{{\bf F_{\Delta\Xi}}}
\newcommand\FYX{{\bf F_{\Xi\Delta}}}
\newcommand\CXX{{\bf C_{\Delta\Delta}}}
\newcommand\CYY{{\bf C_{\Xi\Xi}}}
\newcommand\CXY{{\bf C_{\Delta\Xi}}}
\newcommand\CYX{{\bf C_{\Xi\Delta}}}

\begin{document}

\title{An efficient probe of the cosmological CPT violation}

\author{Gong-Bo Zhao}
\email{gbzhao@nao.cas.cn}
\affiliation{National Astronomy Observatories, Chinese Academy of Science, Beijing, 100012, P. R. China}
\affiliation{Institute of Cosmology \& Gravitation, University of Portsmouth, Portsmouth, PO1 3FX, United Kingdom}

\author{Yuting Wang}

\affiliation{National Astronomy Observatories, Chinese Academy of Science, Beijing, 100012, P. R. China}
\affiliation{Institute of Cosmology \& Gravitation, University of Portsmouth, Portsmouth, PO1 3FX, United Kingdom}

\author{Jun-Qing Xia}

\affiliation{Key Laboratory of Particle Astrophysics, Institute of High Energy Physics, Chinese Academy of Science, P. O. Box 918-3, Beijing 100049, P. R. China}

\author{Mingzhe Li}

\affiliation{Interdisciplinary Center for Theoretical Study, University of Science and Technology of China, Hefei, Anhui 230026, P. R. China}

\author{Xinmin Zhang}
\affiliation{Theory Division, Institute of High Energy Physics, Chinese Academy of Science, P. O. Box 918-4, Beijing 100049, P. R. China}

\begin{abstract}

We develop an efficient method based on the linear regression algorithm to probe the cosmological CPT violation using the CMB polarisation data. We validate this method using simulated CMB data and apply it to recent CMB observations. We find that a combined data sample of BICEP1 and BOOMERanG 2003 favours a nonzero isotropic rotation angle at $2.3\sigma$ confidence level, \ie, $\bar{\alpha}=-3.3\degree\pm1.4\degree$ (68\% CL) with systematics included. 

\end{abstract}

\maketitle

\section{Introduction}

The Charge-Parity-Time (CPT) Reversal invariance is a fundamental symmetry in the standard model (SM) of particle physics, so a CPT-violating signal is a smoking gun of new physics beyond SM. A number of ground-based experiments have been built in order to test this symmetry to high precision. However, no statistically significant signals have been found from these lab-scale experiments so far.   

On the other hand, the CPT symmetry needs to be reexamined on cosmic scales. In the universe, the CPT symmetry can break down in the model with an effective coupling $\mathcal{L}_{\rm int}=\partial_{\mu}f(\phi) J^{\mu}$
between a non-conserved current $J^{\mu}$ of the SM particles and a scalar field $\phi$, which may be identified as the dark energy or the Ricci scalar $R$, and $f$ is an arbitrary function of $\phi$. As an external field, the dynamics of $\phi$ is fixed, therefore the Lorentz and CPT symmetries of the SM particles break down spontaneously. This has been used to interpret the matter-antimatter asymmetry with the current $J^{\mu}$ being the baryon or lepton currents \cite{Li:2001st,Davoudiasl:2004gf,Li:2006ss}. In these models the CPT-violating signal is large at the early stage to generate enough baryon number asymmetry but too small to be detectable by the ground based experiments at present time.

Another interesting phenomenon of the cosmological CPT violation arises if the $J^{\mu}$ is the Chern-Simons current of the electromagnetic field. In this case the polarisation directions of photons rotate when propagating in the space \cite{Carroll:1998zi} \footnote{There are other mechanisms that can break the CPT symmetry, \eg, the circularly-polarised gravitational wave background (GWB) \cite{ss}.}. The rotation angle $\alpha$, which characterises the CPT-violating signal, depends on the difference of $f(\phi)$ between the source and the receiver of photons. It could be large as photons travel across the astronomical and cosmological distances, even though the coupling itself is very small.

The cosmic CPT violation, if exists, could in principle be detected by the observation of the cosmic microwave background radiation (CMB) angular power spectra. Because the rotation of the polarisation vector of the CMB photons can covert part of the E-mode polarisation to B-mode polarisation, the non-zero TB and EB cross correlation, which vanishes in traditional CMB theory, can be induced \cite{Lue:1998mq}. The rotation angle can be probed by observations of the TB and EB power spectra \cite{Feng:2004mq}, and the first constraint on the rotation angle using CMB observations was obtained in \cite{Feng:2006dp}. 

Recent studies along this line have demonstrated that current CMB experiments have the sensitivity to probe the rotation angle at the level of $\mathcal{O}(1\degree)$\cite{Li:2014oia, wmap9, Xia:2012ck, Gruppuso:2011ci, wmap7, Xia:2009ah, bicep1, quad2, wmap5, quad1, Xia:2008si, Xia:2007qs, B03,wavelet}.

As quite a few B-mode CMB experiments are ongoing or being planned, we expect to have high quality observations of the TB and EB spectra, in addition to the TT, EE and TE spectra in the near future. Therefore we need a fast and accurate method to measure the rotation angle using these spectra. Traditional methods, \eg, \cite{Li:2014oia,wmap9,Xia:2012ck,wmap7,Xia:2009ah,wmap5,Xia:2008si,Xia:2007qs,Feng:2006dp}, which are based on the global fitting using the Markov Chain Monte Carlo (MCMC) method, are time consuming and computationally expensive. Alternative methods, which are more efficient, exist, \eg, \cite{Gruppuso:2011ci,quad1,quad2}, but are restricted to the special case where intrinsic BB vanishes. As the hint of a non-zero BB being discovered \cite{bicep2} and investigated \cite{B2K}, an efficient, and more general method including the intrinsic BB spectra are naturally needed. In this paper, we develop such a new method to probe the rotational angle based on a linear regression algorithm. We first validate the method using the simulated CMB data, before applying to recent CMB observations.   

The paper is outlined as follows. In section II, we shall present the methodology, including the tests using mock data. Section III shows the result of the measurement, followed by a discussion and conclusion section.   

\section{Methodology}

In this section, we shall develop the methodology by first presenting the formulism, and then applying it to the simulated CMB data for a validation test. 

\subsection{Formalism}

The rotation angle $\alpha(\hat{\bf n})$ is anisotropic in general and can be decomposed into an isotropic part and an anisotropic one \cite{CCB_theo1}, namely, \be \alpha(\hat{\bf n}) = \bar{\alpha}+\delta\alpha(\hat{\bf n})~. \ee 
The CPT symmetry is violated by the background $\bar{\alpha}$, and the fluctuation $\delta\alpha(\hat{\bf n})$ will bring distortions to the CMB spectra, similar to the lensing effect. The technique based on the quadratic estimator to detect the anisotropies of $\alpha(\hat{\bf n})$ has been developed in Refs. \cite{CCB_theo2,CCB_detect,CCB_test}. If the rotation angle itself has a spectrum, the rotated spectra of CMB can be calculated analytically \cite{CCB_theo1}. 

The general relation between the unrotated and rotated angular power spectra $C_{\ell}$ is \cite{CCB_theo1,LY,Li:2014oia}, \ba\label{eq:Clrot} \cttrot &=&\ctt \nn \\
\cterot &=&A~\cte~{\rm cos}(2\bar{\alpha}) \nn \\
\ctbrot &=& A~\cte~{\rm sin}(2\bar{\alpha}) \nn \\
2\cebrot &=& A^2~f(\ell)~{\rm sin}(4\bar{\alpha}) \nn \\
\ceerot-\cbbrot &=& A^2~f(\ell)~{\rm cos}(4\bar{\alpha})\ea 
Here \be f(\ell)\equiv \sum_{\ell'}\frac{2\ell'+1}{2}(\ceep-\cbbp)\int_{-1}^{1}d_{-22}^{\ell'}(\theta)
d_{-22}^{\ell}(\theta){\rm exp}[-4C^{\alpha}(\theta)]{\rm dcos}(\theta) \ee
where $d^{\ell}_{mn}$ is the Wigner small-$d$ function and the two-point correlation function $C^{\alpha}(\theta)$ is defined as, \be C^{\alpha}(\theta)\equiv \sum_{\ell}\frac{2\ell+1}{4\pi}C^{\alpha\alpha}_{\ell}P_{\ell}({\rm cos}\theta) \ee and $C^{\alpha\alpha}_{\ell}$ is the angular power spectrum of the rotation angle, say, \be \delta\alpha(\hat{\bf n})=\sum _{\ell m} a_{\ell m} Y_{\ell m}(\hat{\bf n}),~~~~\langle a_{\ell m} a^{\ast}_{\ell m} \rangle\equiv C^{\alpha\alpha}_{\ell}\delta_{\ell\ell'}\delta_{m m'} \ee
The overall constant $A$ is related to $C^{\alpha}(0)$ via $A\equiv{\rm exp}[-2C^{\alpha}(0)]$. Note that in the limit that the anisotropic part vanishes, \ie, $\delta\alpha=0$, we have $A=1,~~f(\ell)=\cee-\cbb$ and we get the familiar relation for a constant rotation angle \cite{Feng:2006dp,wmap5,wmap7,wmap9,Xia:2012ck,Xia:2009ah,Xia:2008si,Xia:2007qs}, 

\ba\label{eq:Clrotconst} \cttrot &=&\ctt \nn \\
\cterot &=&\cte~{\rm cos}(2\bar{\alpha}) \nn \\
\ctbrot &=& \cte~{\rm sin}(2\bar{\alpha}) \nn \\
2\cebrot &=& \left(\cee-\cbb\right)~{\rm sin}(4\bar{\alpha}) \nn \\
\ceerot-\cbbrot &=& \left(\cee-\cbb\right)~{\rm cos}(4\bar{\alpha})\ea

From Eq (\ref{eq:Clrot}), it is straightforward to
eliminate the unrotated $C_{\ell}$'s and obtain the following two linear relations between the rotated $C_{\ell}$'s,

\ba\label{eq:est_theo} \ctbrot- {\rm tan}(2\bar{\alpha})~\cterot &=&0 \nn \\
     2\cebrot- {\rm tan}(4\bar{\alpha})~\left(\ceerot-\cbbrot \right)&=&0\ea
Note that these relations generally hold, even if the anisotropy part of the rotation angle exists (since it drops out), and they are cosmology-independent simply because the theoretical, unrotated $C_{\ell}$'s do not show up here. This, in principle, allows the estimation of $\bar{\alpha}$ directly from data. In what follows, we shall develop a new method based on the linear regression to estimate $\bar{\alpha}$, validate it using simulated data and apply it to the real CMB observations. 

The observed angular power spectra, $C_{\ell}^{\rm obs}$, are the rotated ones with measurement uncertainties. Therefore, the left hand side of Eq (\ref{eq:est_theo}) do not strictly vanish when $C_{\ell}^{\rm rot}$ is replaced by $C_{\ell}^{\rm obs}$, but with residues $\Delta_{\ell}$ and $\Xi_{\ell}$, namely, \ba \Delta_{\ell}&\equiv&\ctbobs- {\rm tan}(2\bar{\alpha})~\cteobs \nn  \\
\Xi_{\ell}&\equiv&2\cebobs- {\rm tan}(4\bar{\alpha})~\left(\ceeobs-\cbbobs \right)\ea 

It is a typical linear regression problem to find $\bar{\alpha}$ from the above relation, and similar problems have been studied extensively in other fields of astronomy \cite{linreg1,linreg2}. Here we briefly review these methods. 

For a linear regression problem\footnote{We set the intercept to zero here since it does not apply for our concerned problem.}, namely, \be y_i=\beta x_i,  \ee where the $(x_i, y_i)$ pair denotes the $i$th measurement with errors and the slope $\beta$ is the free parameter to be determined. The developed linear regression methods include ${\tt OLS}$ (Ordinary Least Squares), ${\tt BCES}$ \cite{BCES}, ${\tt FITEXY}$ \cite{FITEXY, FITEXYmod}, ${\tt MLE}$ (Maximum Likelihood Estimates) \cite{MLE} and the sophisticated full Bayesian Gaussian mixing method \cite{linreg2}. Ref. \cite{linreg1} extensively compared the behaviour of all these methods based on the simulated data obeying various distributions, and concluded that the ${\tt BCES}$, ${\tt FITEXY}$, and the Bayesian methods work equally well. Here we adopt the ${\tt FITEXY}$ model due to its robustness and simplicity.

The {\tt FITEXY} method was originally proposed in \cite{FITEXY} and modified by \cite{FITEXYmod}. This method determines $\beta$ by minimising the following $\chi^2$ (we ignore the correlations among data points for the moment), 
\be\label{eq:chi2} \chi^2=\sum_{i=1}^N\frac{\left(y_i-\beta x_i\right)^2}{\sigma_{y,i}^2+\beta^2\sigma_{x,i}^2+\sigma_{\rm int}^2} \ee 
Here $\sigma_{\rm int}$ accounts for the intrinsic scatter in the measurement of $y$, and its value is to be determined by iteration such that the reduced $\chi^2$ for the best fit model equals one, \ie, $\chi^2_{\rm red}\equiv\chi_{\rm BF}^2/(N-1)=1$. This is to guarantee that the dataset is never under-fitted in order to avoid the bias. As shown in Ref. \cite{linreg1}, the ${\tt OLS}$ method (\ie, when the intrinsic scatter is not included) can lead to severely biased result. 

In general cases where the measurements are correlated, \ie, \be {\rm cov}(x_i,x_j)\ne0;~~~~ {\rm cov}(y_i,y_j)\ne0;~~~~{\rm cov}(x_i,y_j)\ne0, \ee we generalise Eq (\ref{eq:chi2}) to,  \be \chi^2=\left(\Delta~\Xi\right)  \left(\begin{array}{cc}\FXX & \FXY \\ \FYX & \FYY \end{array}\right)  \left(\begin{array}{c}\Delta \\ \Xi \end{array}\right).\ee where \be {\bf F^{-1}}\equiv{\bf C}=\left(\begin{array}{cc}\CXX & \CXY \\ \CYX & \CYY \end{array}\right) \ee and 
\ba \label{eq:C} \left({\bf C_{\Delta\Delta}}\right)_{\ell\ell'}&\equiv& {\rm cov}(\Delta_{\ell}, \Delta_{\ell'}), \nn \\
\left({\bf C_{\Xi\Xi}}\right)_{\ell\ell'}&\equiv& {\rm cov}(\Xi_{\ell}, \Xi_{\ell'}), \nn \\
\left({\bf C_{\Delta\Xi}}\right)_{\ell\ell'}&\equiv& {\rm cov}(\Delta_{\ell}, \Xi_{\ell'}),\nn \\
\left({\bf C_{\Xi\Delta}}\right)_{\ell\ell'}&\equiv& {\rm cov}(\Xi_{\ell},\Delta_{\ell'}). \ea 
Note that $\bf C_{\Delta\Xi}$ and $\bf C_{\Xi\Delta}$ are not symmetric simply because ${\rm cov}(\Delta_{\ell}, \Xi_{\ell'})\ne{\rm cov}(\Delta_{\ell'}, \Xi_{\ell})$ but ${\bf C}$ is symmetric. These covariance matrices can be directly calculated given the full data covariance matrices, namely \footnote{We drop the superscript `obs' for brevity.},

\ba {\rm cov} \left(\Delta_{\ell},~\Delta_{\ell'}\right)&=&{\rm cov} \left[\ctb-{\rm tan}(2\alpha)~\cte,~~\ctbp-{\rm tan}(2\alpha)~\ctep\right] \nn \\ 
&=&  {\rm cov} \left[\ctb,~~\ctbp\right] -{\rm tan}(2\alpha)~\left\{{\rm cov}\left[\cte,~~\ctbp\right]+ {\rm cov}\left[\ctb,~~\ctep\right]\right\} \nn \\
&&+~{\rm tan}^2(2\alpha)~{\rm cov}\left[\cte,~~\ctep\right]  \nn \\ \nn \\
{\rm cov} \left(\Xi_{\ell},~~\Xi_{\ell'}\right)&=&{\rm cov} \left[2\ceb-{\rm tan}(4\alpha)~\left(\cee-\cbb\right),~~2\cebp-{\rm tan}(4\alpha)~\left(\ceep-\cbbp\right)\right] \nn \\ 
&=& ~4~{\rm cov} \left[\ceb,~\cebp\right] \nn \\
&&-2~{\rm tan}(4\alpha)~\left\{{\rm cov}\left[\ceb,~~\ceep\right]-{\rm cov}\left[\ceb,~~\cbbp\right]+{\rm cov}\left[\cee,~~\cebp\right]-{\rm cov}\left[\cbb,~~\cebp\right] \right\}\nn \\
&&+~{\rm tan}^2(4\alpha)~\left\{{\rm cov}\left[\cee,~~\ceep\right]-{\rm cov}\left[\cee,~~\cbbp\right]-{\rm cov}\left[\cbb,~~\ceep\right]+{\rm cov}\left[\cbb,~~\cbbp\right] \right\} 
\nn \\ \nn \\
{\rm cov} \left(\Delta_{\ell},~\Xi_{\ell'}\right)&=&{\rm cov} \left[\ctb-{\rm tan}(2\alpha)~\cte,~~2\cebp-{\rm tan}(4\alpha)~\left(\ceep-\cbbp\right)\right] \nn \\
&=&2~{\rm cov} \left[\ctb,~\cebp\right] - {\rm tan}~(4\alpha)\left\{{\rm cov} \left[\ctb,~\ceep\right]- {\rm cov} \left[\ctb,~\cbbp\right]\right\}- 2~{\rm tan}~(2\alpha){\rm cov} \left[\cte,~\cebp\right] \nn \\
&&+~ {\rm tan}~(2\alpha)~{\rm tan}~(4\alpha)\left\{{\rm cov} \left[\cte,~\ceep\right]- {\rm cov} \left[\cte,~\cbbp\right]\right\} \nn \\
\nn \\
{\rm cov} \left(\Xi_{\ell},~\Delta_{\ell'}\right)&=&{\rm cov} \left(\Delta_{\ell},~\Xi_{\ell'}\right)(\ell\leftrightarrow	\ell')\ea

So given the binned measurements of $\ctb$ and $\cte$, or $\cee,\cbb$ and $\ceb$, the rotation angle can be determined without degeneracy with any other cosmological parameters, which is one of the advantages of this method.

\subsection{Tests using mock data}

\begin{figure}
\centering
{\includegraphics[scale=0.25]{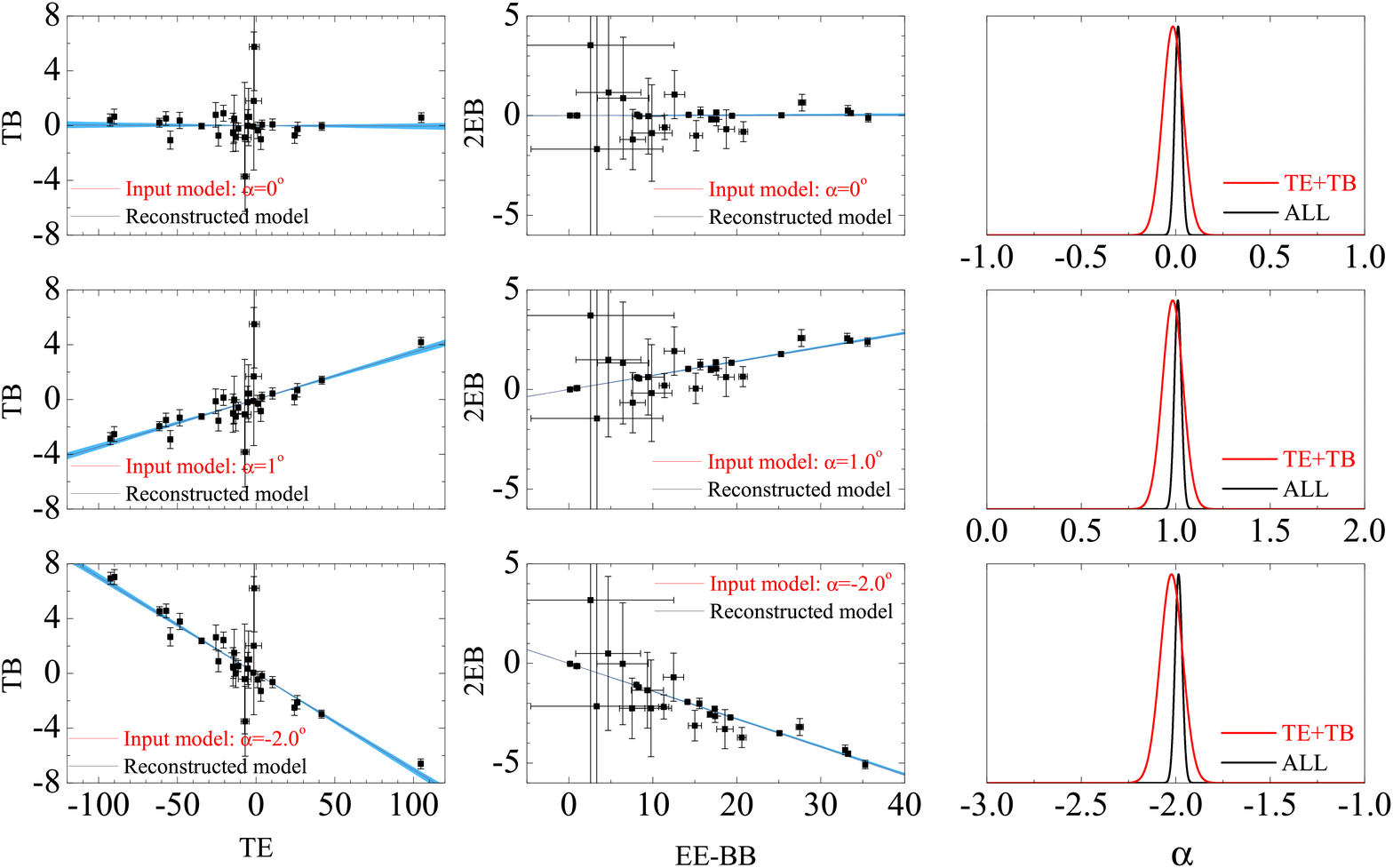}}
\caption{The validation test result using mock data. Each row of panels shows the result for a given fiducial model of rotation angle (top: $\alpha=0\degree$; middle: $\alpha=1\degree$; bottom: $\alpha=-2\degree$). In each row, the left (middle) panel shows the linear regression result using mock data of TB and TE (2EB and EE-BB), where the data points with error bars are mock CMB data; the red (black) line shows the input (reconstructed) model. The blue error band shows the 68\% CL uncertainty of the reconstruction. The panels in the right column show the one-dimensional posterior distribution of $\alpha$ derived from the linear regression method for three fiducial models. The red and black curves show the result using TE+TB and all the CMB data (including EE, BB and EB) respectively.}
\label{fig:mock}
\end{figure}

To test the validity of this method, we generate mock CMB data following \cite{Li:2014oia} and assume a Planck sensitivity summarised in Table 1 therein. The noise power spectra is then estimated as,
\ba N^T_{\ell,{\rm c}}&=&(\Delta_T \theta_{\rm FWHM, c})^2 {\rm exp}\left[\frac{\ell(\ell+1)\theta^2_{\rm FWHM, c}}{8~{\rm ln}2}  \right] \nn \\
N^P_{\ell,{\rm c}}&=&(\Delta_P \theta_{\rm FWHM, c})^2 {\rm exp}\left[\frac{\ell(\ell+1)\theta^2_{\rm FWHM, c}}{8~{\rm ln}2}  \right] \ea
where $T$ and $P$ denote the `Temperature' and `Polarisation' respectively, and $\theta_{\rm FWHM, c}$ is the Full width at half maximum (FWHM) of the angular resolution for a given frequency channel c. The combined noise from all channels is then, \ba N^T_{\ell}&=& \left[\sum_{c} \left(N^T_{\ell,{\rm c}}   \right)^{-1} \right]^{-1} \nn \\
N^P_{\ell}&=& \left[\sum_{c} \left(N^P_{\ell,{\rm c}}   \right)^{-1} \right]^{-1}\ea The observed dimensionless $C_{\ell}$'s (without the $\ell(\ell+1)/2\pi$ factor) are, \be C_{\ell}^{\rm obs} = C_{\ell}^{\rm theo}+N_{\ell} \ee
We randomly displace the data points by the corresponding 1-$\sigma$ error. For this mock test, we choose a Planck best-fit cosmology with the tensor-to-scalar ratio $r=0.1$, and with several fiducial values of the rotation angle, say, $\bar{\alpha}=0\degree, 1\degree, -2\degree$. The mock data points are shown in the left and middle panels in Fig \ref{fig:mock}. One can easily identify the slope in this plot visually, which is useful for a quick consistency check. For example, if the slope in the TB-TE data has the opposite sign in the EE-BB-EB plot, it might suggest an inconsistency in the dataset itself. We then apply the {\tt FITEXY} method to measure the slope between TB and TE, and between 2EB and (EE-BB) (we further drop the symbol of $C_{\ell}$ for brevity). The measurements (best-fit value and 68\% CL uncertainty) are over-plotted with data points in Fig \ref{fig:mock}. Given the measurement of the slope, we obtain the constraint on $\bar{\alpha}$, which is shown in the right panels of Fig \ref{fig:mock} and in the top part of Table I. As shown, our method works very well in all three cases, namely, the input models are accurately reconstructed. 

\begin{table}[htdp]
\begin{center}
\begin{tabular}{c|c|c|c}

\hline\hline 
Fiducial model for mock data & {TE+TB} & {EE+BB+EB} &{ALL} \\
\hline
$\alpha_{\rm fid}=0\degree$ & $-0.015\degree\pm0.054\degree$ & $0.016\degree\pm0.020\degree$ & $0.013\degree\pm0.018\degree$   \\
$\alpha_{\rm fid}=1\degree$ & $0.984\degree\pm0.054\degree$ & $1.015\degree\pm0.020\degree$ & $1.011\degree\pm0.019\degree$  \\
$\alpha_{\rm fid}=-2\degree$ & $-2.020\degree\pm0.058\degree$ & $-1.981\degree\pm0.020\degree$ & $-1.985\degree\pm0.019\degree$  \\
\hline
\hline

\end{tabular}
\end{center}
\label{tab:mock}
\caption{The measurement using the mock data (best fit value with 68\% CL uncertainty) on the isotropic rotation angle. }
\end{table}%

\section{Result}

\begin{figure}[htdp]
\centering
{\includegraphics[scale=0.5]{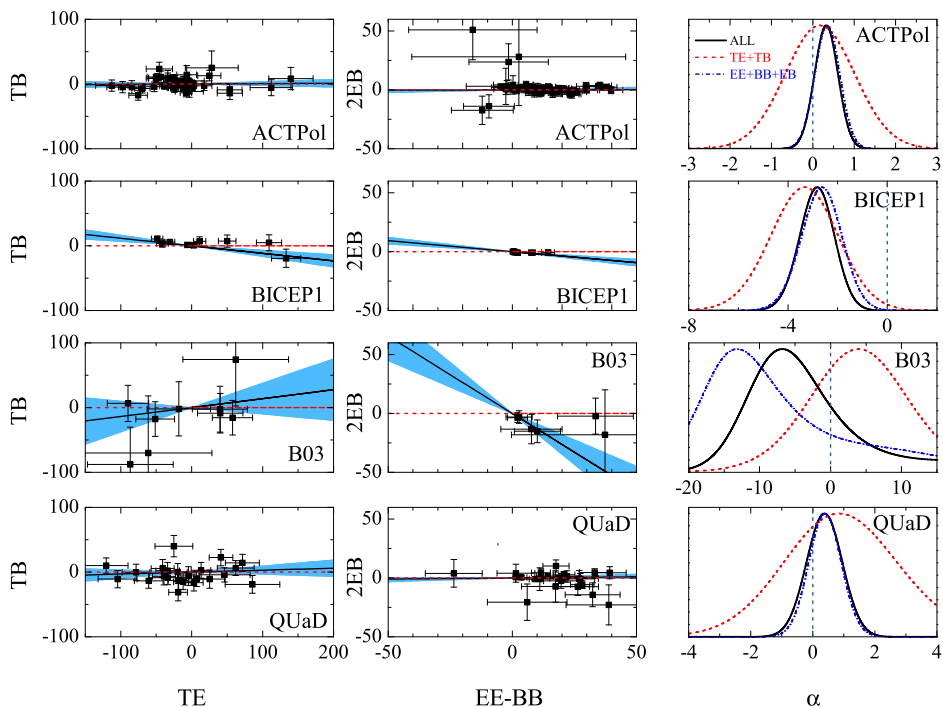}}
\caption{The $\alpha$ measurement result using recent CMB polarisation observations. In each row, the left (middle) panel shows the linear regression result using the data of TB and TE (2EB and EE-BB), where the data points with error bars are CMB data; the red (black) line shows the input (reconstructed) model. The blue error band shows the 68\% CL uncertainty of the reconstruction. The horizontal red dashed lines shows TB = EB = 0. The panels in the right column show the one-dimensional posterior distribution of $\alpha$. The red dashed, blue dash-dotted and black solid curves show the result using TE+TB, EE+BB+EB and combined data respectively. The vertical dashed lines show $\alpha=0$ to guide eyes.}
\label{fig:realdata}
\end{figure}

\begin{figure}[htdp]
\centering
{\includegraphics[scale=0.25]{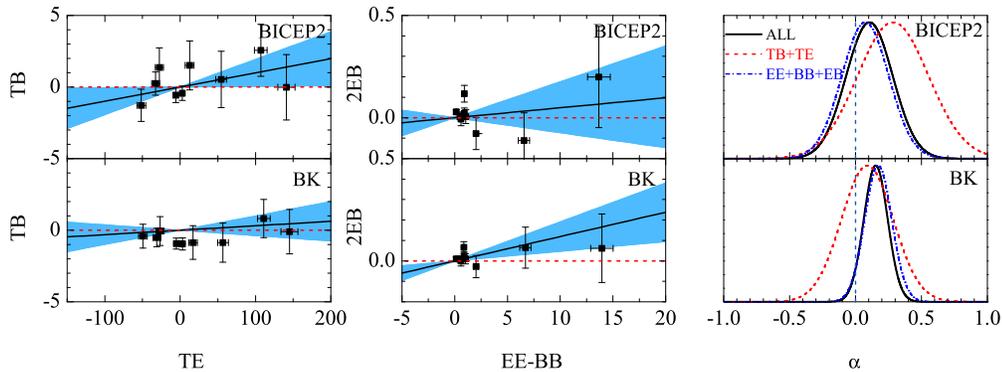}}
\caption{Same as Fig 2 but for the most recent BICEP2 and BICEP2/Keck data.}
\label{fig:BK}
\end{figure}

In this section, we apply our method to recent CMB measurements, including ACTPol \cite{ACTPol}, BICEP1 \cite{bicep1}, BOOMERanG 2003 (B03) \cite{B03}, QUaD \cite{quad1,quad2}, and we show the result in Figs. \ref{fig:realdata}, \ref{fig:BK} and in Table II. From Fig \ref{fig:realdata}, we can see that the the constraints from ACTPol and QUaD are consistent with zero rotation, but the slopes for the datasets of BICEP1 and B03 are nonzero at more than 68\% confidence level. Our measurement shows, \ba {\rm B03:}&& \bar{\alpha}=3.7\degree\pm6.8\degree~~({\rm TE+TB});~~\bar{\alpha}=-13.1\degree\pm4.6\degree~~({\rm EE+BB+EB});~~\bar{\alpha}=-8.1\degree\pm4.5\degree~~({\rm ALL}) \nn \\ 
{\rm BICEP1:}&& \bar{\alpha}=-3.6\degree\pm1.2\degree~~({\rm TE+TB});~~\bar{\alpha}=-2.6\degree\pm0.8\degree~~({\rm EE+BB+EB});~~\bar{\alpha}=-2.9\degree\pm0.7\degree~~({\rm ALL})
\ea To be more conservative, we follow \cite{Pagano,Xia:2012ck,Li:2014oia} to marginalise over another rotation angle $\eta$ with the following Gaussian prior to account for the possible systematics, namely, \be  \eta_{\rm B03}=-0.9\degree\pm0.7\degree;~~\eta_{\rm BICEP1}=0.0\degree\pm1.3\degree\ee With the systematics included, the constraint is diluted to, \be {\rm B03:}~~\bar{\alpha}=-7.2\degree\pm4.6\degree~~({\rm ALL+sys.});~~{\rm BICEP1:}~~\bar{\alpha}=-2.9\degree\pm1.5\degree~~({\rm ALL+sys.}) \ee which is consistent with the published result in \cite{Li:2014oia}. The combined dataset gives, \be  {\rm B03+BICEP1:}~~\bar{\alpha}=-3.3\degree\pm1.4\degree ({\rm ALL+sys.}) \ee This shows a 2.3 $\sigma$ signal of a non-zero rotation angle. 

This signal can be visually identified in the BICEP1 and B03 panels of Fig \ref{fig:realdata}. In both TB and EB panels of BICEP1, we can see a consistent negative slope at more than $1\sigma$ level. However, the TB panel of B03 does not show any significant slope, but its EB panel shows a negative slope at $~3\sigma$ level. The origin of this slope needs to be further investigated by future CMB experiments.     

Recently the BICEP2 collaboration published the new measurement of the CMB polarisation data \cite{bicep2}, and a joint analysis between BICEP2 and Keck observation (BK) has been performed \cite{B2K}. These are the most precision measurement of the CMB polarisation. But unfortunately these data cannot be used to constrain the rotation angle because they were already self-calibrated \cite{SC}, so that any nonzero rotation angle, if exists, has been removed from the maps. But as a consistency test, we apply our method on these data, and show the result in Fig \ref{fig:BK} and in the lower part of Table II. As shown, the constraint from the BICEP2 data is very consistent with $\alpha=0$, but the BK dataset shows a slight preference to a non-zero rotation angle, although the significance is low ($1.7\sigma$). This signal shows up when using the EE, BB and EB spectra, which gives a $1.7\sigma$ signal without using TE and TB. This might suggest that the self-calibration operation performed for the BK data is not as complete as that for the BICEP2 data.   

\begin{figure}
\centering
{\includegraphics[scale=0.5]{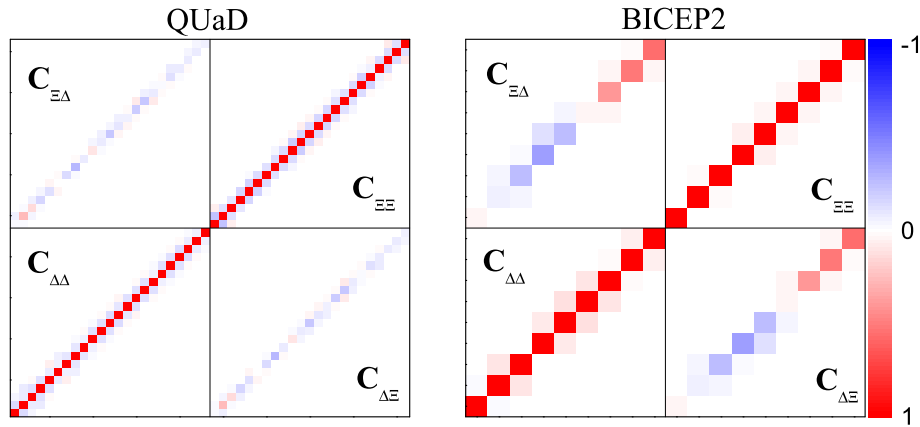}}
\caption{The correlation matrix for the best fit model using the QUaD and BICEP2 polarisation data respectively. The upper right and lower left blocks show the correlation coefficients between $\ell$ bins for the same kind of data ($\bf C_{\Xi\Xi}$ and $\bf C_{\Delta\Delta}$ respectively), while the rest two blocks are for the correlation between $\ell$ bins for different kinds of data ($\bf C_{\Xi\Delta}$ and $\bf C_{\Delta\Xi}$ respectively). As shown, the correlation is only visible between neighbouring bins for the same kind of data, or between different kinds of data but for the same bins.}
\label{fig:corr}
\end{figure}

We show the correlation matrix, which is the rescaled covariance matrix ${\bf C}$ defined in Eq (\ref{eq:C}),  for the best fit rotation angle using QUaD and BICEP2 data in Fig \ref{fig:corr} \footnote{Note that the correlation matrix depends on $\alpha$, so we need to specify the value of rotation angle for the illustration. }. In each panel, the upper right and lower left blocks show the correlation coefficients between $\ell$ bins for the same kind of data ($\bf C_{\Xi\Xi}$ and $\bf C_{\Delta\Delta}$ respectively), while the rest two blocks are for the correlation between $\ell$ bins for different kinds of data ($\bf C_{\Xi\Delta}$ and $\bf C_{\Delta\Xi}$ respectively). As shown, the correlation is only non-negligible between neighbouring bins for the same kind of data, or between different kinds of data but for the same bins, which is expected. 

\begin{figure}
\centering
{\includegraphics[scale=0.4]{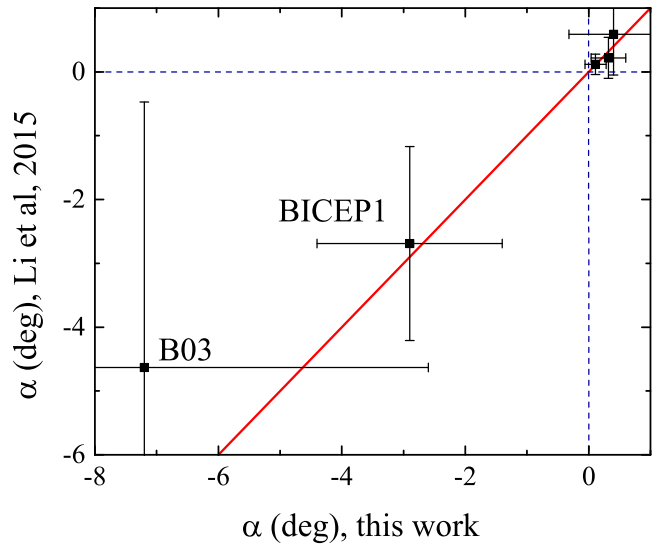}}
\caption{A visual comparison of the constraint on the rotation angle between this work and \cite{Li:2014oia}. The red line illustrates the perfect consistency, and the horizontal and vertical dashed lines show $\alpha=0$.}
\label{fig:compare}
\end{figure}

This constraint on the rotation angle using our new, efficient method is largely consistent with those using the MCMC global fitting methods, as recently performed in \cite{Li:2014oia}. Fig \ref{fig:compare} shows the consistency, namely, all the measurements lay on the diagonal red line within $1\sigma$ error bars. The two measurements using B03 and BICEP1 respectively deviate from $\alpha=0$ noticeably, which is discovered by both works.

\begin{table*}[htdp]
\begin{center}
\begin{tabular}{c|c|c|c|c}

\hline\hline 
Experiments & {TE+TB} & {EE+BB+EB} &{ALL (ALL+systematics)}&Published Result \\
\hline

ACTPol \cite{ACTPol}       & $0.17\degree\pm0.84\degree$  &$0.34\degree\pm0.30\degree$ &$0.32\degree\pm0.28\degree$  &  $0.22\degree\pm0.32\degree$ \cite{ACTPolresult} \\
B03  \cite{B03}            & $3.7\degree\pm6.8\degree$  &$-13.1{\degree}\pm4.6\degree$  &$-8.1{\degree}\pm{4.5\degree}(-7.2\degree\pm4.6\degree )$ & $-4.63\degree\pm4.16\degree$ \cite{Li:2014oia}\\
BICEP1\cite{bicep1}       & $-3.6\degree\pm1.2\degree$  &$-2.6\degree\pm0.8\degree$  &$-2.9\degree\pm0.7\degree (-2.9\degree\pm1.5\degree)$ &  $-2.69\degree\pm1.52\degree$ \cite{Li:2014oia}\\
QUaD\cite{quad1,quad2}          & $0.86\degree\pm1.79\degree$  &$0.41\degree\pm0.52\degree$  &$0.40\degree\pm0.52\degree (0.40\degree\pm0.72\degree)$ & $0.59\degree\pm0.64\degree$ \cite{Li:2014oia}\\
\hline
BICEP2\cite{bicep2}       & $0.28\degree\pm0.27\degree$  &$0.08\degree\pm0.17\degree$ &$0.11\degree\pm0.17\degree$  & $0.12\degree\pm0.16\degree$ \cite{Li:2014oia}\\
BK\cite{B2K}       & $0.09\degree\pm0.19\degree$  &$0.17\degree\pm0.10\degree$  &$0.15\degree\pm0.09\degree$  & \\

\hline\hline 

\end{tabular}
\end{center}
\label{tab:real}
\caption{The measurement (best fit value with 68\% CL uncertainty) on the isotropic rotation angle using different experiments. The numbers quoted in the parenthesis in the right column are those corrected with systematics.}
\end{table*}%

\section{Conclusion and discussions}

Constraining the rotation angle of the polarisation vector of the CMB photons is an effective way to test the CPT symmetry on cosmological scales. Given that the CMB polarisation measurements are getting more and more accurate, and that several B-mode experiments are ongoing or being planned, it is timely to develop accurate, and efficient methods to probe the rotation angle. 

In this paper, we have developed a new method for this purpose based on the linear regression algorithm. Compared to previously established methods, our new method is computationally efficient (it takes several minutes on a laptop), accurate, intuitively transparent, and more generally applicable. This new method has successfully passed the validation tests using mock CMB data before applied to CMB observations, including ACTPol, B03, BICEP1, QUaD, as well as BICEP2 and BK. We find that the B03 and BICEP1 samples prefer a nonzero rotation angle even when the systematics are included, and a combination of these two surveys gives $\bar{\alpha}=-3.3\degree\pm1.4\degree$, which is a $2.3\sigma$ signal. However, ACTPol and QUaD support a zero rotation angle. BICEP2 and BK provide the most stringent constraint on the rotation angle, namely, the uncertainty reaches $\sim0.1\degree$ level, but unfortunately they cannot be used to probe for the rotation angle since the datasets have been self-calibrated. 

With the high-quality polarisation data available in the near future, we would be able to constrain the anisotropic rotation angle $\delta\alpha$, as attempted by \cite{CCB_theo1,CCB_theo2,CCB_detect,CCB_test,LY,Li:2014oia}. Since $\delta\alpha$ is degenerate with $\bar{\alpha}$ using traditional methods, our method is efficient to break the degeneracy since $\bar{\alpha}$ is not degenerate with any other cosmological parameters, including $\delta\alpha$, in our prescription. An extension of this work is to develop a new algorithm to probe the anisotropic rotation angle using the regression method and a principle component analysis, which is left for a future study.

\acknowledgements

We thank Andrei Frolov and Levon Pogosian for insightful discussions, and Brian Keating for the correspondence regarding the BICEP2 and Keck dataset. GBZ are JQX are supported by the 1000
Young Talents program in China. GBZ, YW, JQX and XZ are supported by the Strategic Priority Research Program
``The Emergence of Cosmological Structures" of the Chinese
Academy of Sciences, Grant No. XDB09000000. GBZ is supported by the 973 Program
grant No. 2013CB837900, NSFC grant No. 11261140641,
and CAS grant No. KJZD-EW-T01. YW is supported by the NSFC grant No. 11403034 and the China Postdoctoral Science Foundation Grant No. 2014M550091. 
ML is supported in part by NSFC grant No. 11422543 and the Fundamental Research Funds for the Central Universities. XZ is supported in part by NSFC grants 11121092, 11033005 and 11375202.

\end{document}